\documentclass[twocolumn,10pt]{asme2e}

\usepackage{xcolor}
\usepackage{mathtools}
\usepackage{gensymb}

\confshortname{DSCC 2019}
\conffullname{the ASME 2019 \\ Dynamic Systems and Control Conference}

\confdate{9-11}
\confmonth{October}
\confyear{2019}
\confcity{Park City, Utah}
\confcountry{USA}

\papernum{DSCC2019-9242}

\title{Modeling height profile for Drop-on-demand print\\ of UV curable ink}

\author{Yumeng Wu
    \affiliation{
	School of Mechanical Engineering\\
	Purdue University\\
	West Lafayette, Indiana 47907\\
    Email: wu350@purdue.edu
    }	
}

\author{George Chiu\thanks{Address all correspondence to this author.} 
    \affiliation{
	School of Mechanical Engineering\\
	Purdue University\\
	West Lafayette, Indiana 47907\\
    Email: gchiu@purdue.edu
    }
}

\begin{document}
\maketitle
\begin{abstract}
This paper proposes a height profile model for drop-on-demand printing of UV curable ink. 
Existing models include superposition of single drops, numerical models, and graphic-based model.
They are either too complicated or over simplified.
Graphic model intends to find a sweet spot in between, however, accuracy is marginally improved from superposition model while it demands more computation.
The proposed model aims to achieve the same as graphic model by introducing volume and area propagation matrices to reflect the localized ink flow from higher location to the lower, while avoiding the detailed physics behind it.
This model assumes a constant volume and area propagation of subsequent drop due to height profile difference.
It is validated with experiments on single drop, 2-drop and 3-drop line printing.
Stability of this model is analyzed..
Using root mean square (RMS) error as benchmark, proposed model achieves 6.6\% along the center row and 7.4\% overall, better than existing models.
\end{abstract}

\begin{nomenclature}
\entry{$a_{i j}$}{Area covered by a single drop at ($i,j$).}
\entry{$v_{i j}$}{Drop volume at cell ($i,j$).}
\entry{$h_{i j}$}{Average height of a single drop at cell ($i,j$).}
\entry{$a_{i j, k}$}{Area coverage at cell ($i,j$) after the $k^{th}$ drop is deposited.}
\entry{$v_{i j, k}$}{Drop volume at cell ($i,j$) after the $k^{th}$ drop is deposited.}
\entry{$h_{i j, k}$}{Average height at cell ($i,j$) after the $k^{th}$ drop is deposited.}
\entry{$\Delta v_{i j}$}{Volume change at cell ($i, j$).}
\entry{$\Delta a_{i j}$}{Area change at cell ($i, j$).}
\entry{$A_{j}[0]$}{Area matrix in the new area of interest before additional drop is deposited  at cell (1,$j$).}
\entry{$V_{j}[0]$}{Volume matrix in the new area of interest before additional drop is deposited  at cell (1,$j$).}
\entry{$A_{j}[1]$}{Area matrix in the new area of interest after additional drop is deposited  at cell (1,$j$).}
\entry{$V_{j}[1]$}{Volume matrix in the new area of interest after additional drop is deposited  at cell (1,$j$).}
\entry{$H_{j}[1]$}{Height matrix in the new area of interest after additional drop is deposited  at cell (1,$j$).}
\entry{$d$}{Nominal pitch between two drops.}
\entry{$r$}{Radius of single drop's footprint on substrate.}
\end{nomenclature}

\section{\MakeUppercase{Introduction}}

3D printing, or additive manufacturing (AM), has gained traction in interests and applications. 
Comparing to traditional subtractive manufacturing, e.g. CNC machining, the benefits of AM include less waste of materials, ability to manufacture more complex geometry and lower cost on small to median batch production \cite{ford2016additive}. 

Among different additive manufacturing processes \cite{standard2012iso}, material jetting is unique for its similarity with traditional ink-jet printing.
Two fundamental categories of ink-jet printing are continuous jetting and drop-on-demand (DoD) \cite{le1998progress}.
In DoD inkjet, drops are formed by either squeezing with piezoelectric ink-jet or heating with thermal ink-jet.
While traditional ink-jet printing can dispense a wide range of materials, from color inks to polymers and nanoparticles \cite{sirringhaus2003inkjet}, material jetting requires inks that can be solidified under prescribed  conditions.
Among such materials, photosensitive inks, which solidifies under light of certain wavelengths, are commonly used.
Commercial printers, such as Acuity EY series from FUJIFILM, HP Scitex FB500, use UV curable inks.
Typically, cured drop diameters range from 50 to 500 $\mu m$, depending on the nozzle size \cite{cooley2002applicatons}.
With decades of studies on ink-jet printing, many approaches have been proposed to improve print quality \cite{Agar2005, Li2004}. 
Majority of these approaches focus on color and image quality. 
Mimaki shows that material jetting is capable of doing additive manufacturing in full color.
However, the height resolution is lower than that of color.
Adapting optimization on color and image quality to geometry, especially height profile, of products made by material jetting can be effective.
Such optimization requires a height profile model to balance between accuracy and computation.

For DoD inkjet print of UV curable ink, a single drop is commonly modeled as a spherical cap \cite{Doumanidis2000}. 
Treating multiple drops as superposition of single drop is commonly applied, but leads to large errors, because many key physical properties are ignored.
There are other numerical models considering many parameters \cite{choi2017numerical,gunjal2003experimental}.
Guo \cite{Guo2018} introduced a model for the purpose of control, however, the graph-based dynamic model propagates height directly does not guarantee conservation of volume.
Moreover, its root mean square (RMS) errors are often greater than 10\%, which is on par with simple superposition model. 

In this paper, a height profile model is developed to reflect the physical movement of inks on non-porous surface when the subsequent drop is deposited next to a cured drop. 
This model calculates height indirectly from volume distribution and area propagation.
The remaining of the paper is organized as follow. the height profile model is introduced in Section \ref{sec:model}.
Propagation model is discussed in Section \ref{sec:model}.
Experimental validation is included in Section \ref{sec:exp_valid}.
Section \ref{sec:conclusion} is the conclusion.

\section{\MakeUppercase{Single Drop Height Profile Model}}
\label{sec:single}
\begin{figure}
	\centering
	\includegraphics[width=\linewidth]{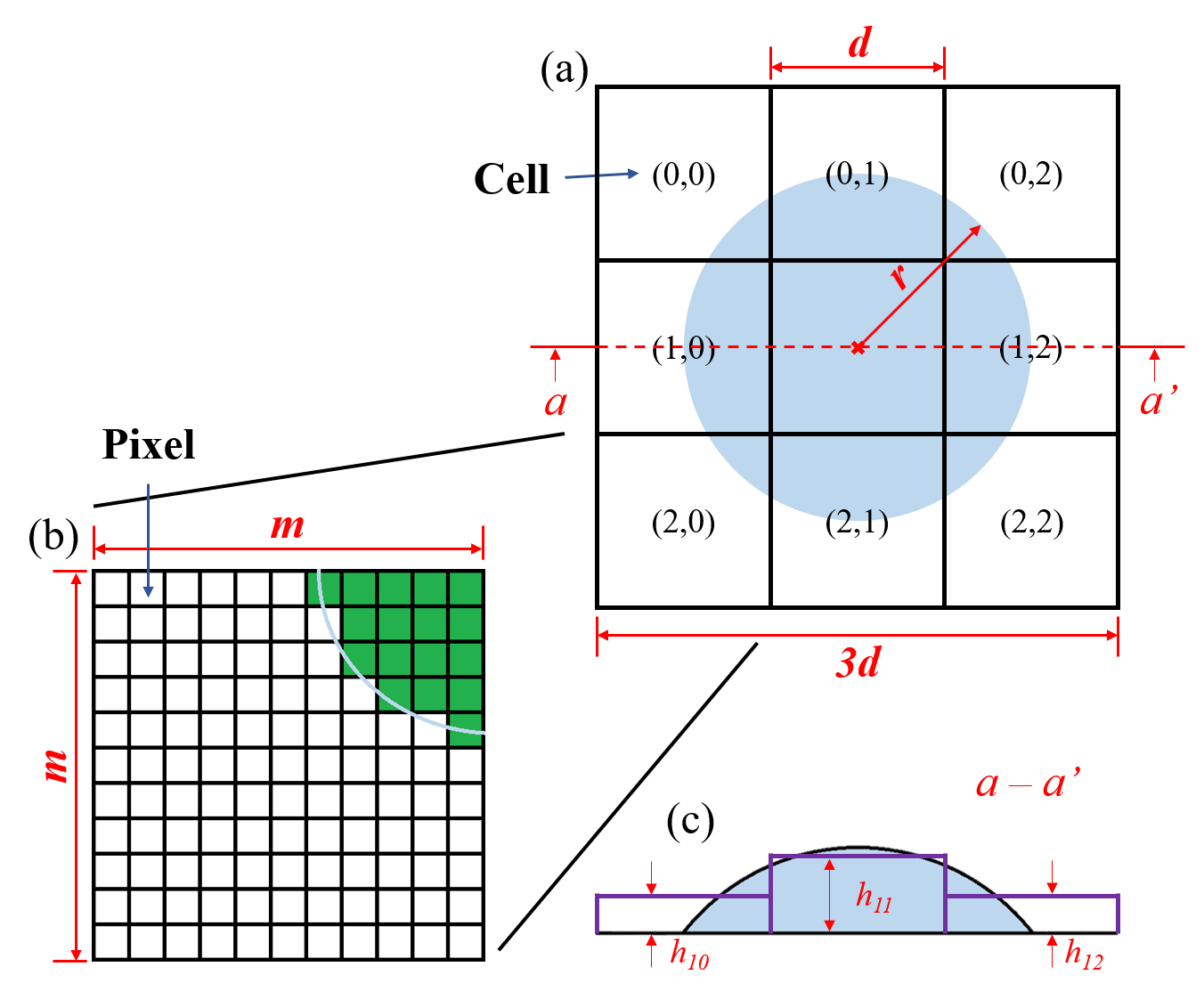}
	\caption{(a): Single drop with radius $r$ on $3 \times 3$ cell. \hspace{\textwidth} 
	(b): Zoomed-in section of cell (2,0) in (a), illustrating how to count area covered by inks.
	(c): Height profile comparison of center row. Proposed height profile is in purple, while spherical cap height profile is in light blue.}
	\label{fig:single}
\end{figure}

Figure \ref{fig:single} illustrates a single drop height profile model and defines notations.
Figure \ref{fig:single}.(a) shows $3 \times 3$ square cells, where a drop of material can be deposited at the center of each cell.
The size of each cell is $d \times d$, where $d$ is the pitch between two adjacent drops. 
Assuming a single drop deposited on a flat non-porous surface  forms a spherical cap with radius $r$, if $0.5 d \le r <1.5 d$, the drop covers the $3 \times 3$ cells, see the shaded area in Fig. \ref{fig:single}(a). 
This range of pitch distance is used to make enough overlap between adjacent drops to ensure print quality. 
Each cell is indexed by ($i,j$), where $i$ is the row index and $j$ is the column index.
Both $i$ and $j$ start from 0.

The height profile of a single drop is denoted by a $3 \times 3$ matrix,
\begin{equation}
\label{eq:h_k}
    H = \left[ \begin{array}{ccc}
       h_{0 0} &  h_{0 1} &  h_{0 2}\\
       h_{0 0} &  h_{1 1} &  h_{1 2}\\
       h_{0 0} &  h_{2 1} &  h_{2 2}  
\end{array} \right],
\end{equation}
where $h_{ij}$ denotes the average height of the drop in cell ($i,j$),
which is computed from the ratio between the drop volume within the cell and the area covered by the drop within the cell covered by the drop.
Figure \ref{fig:single}.(c) illustrates the height profiles of cells (1,0), (1,1) and (1,2). 
Area in light blue represents the height profile from the spherical cap model.
Horizontal purple lines represent the proposed height profile.

To compute the height profile, a high-resolution optical profilometer is used to measure the drop profile with fine spatial resolution.
Figure \ref{fig:single}.(b) illustrates the optical profilometer image for cell (2,0) of Fig. \ref{fig:single}.(a), where each cell image is represented by $m \times m$ pixels.
The area under each pixel ($a_c$) is $\frac{d^2}{m^2}$.
In Fig. \ref{fig:single}.(b), the area covered by the drop is shown in green.
For each pixel, the height $h_{i_m j_m}$ is non-zero if it is covered by the drop and is zero if it is not covered by the drop, where $i_m,j_m=0,\cdots,m-1$ are pixel indices within a cell.
Assuming $M$ pixels within cell ($i,j$) are covered by the drop, the area occupied by the drop $a_{i j} = M a_c$. 
Similar to the height profile, the area covered by a single drop can be denoted by 
\begin{equation}
\centering
\label{eq:a_single}
    A = \left[ \begin{array}{ccc}
       a_{0 0} &  a_{0 1} &  a_{0 2}\\
       a_{0 0} &  a_{1 1} &  a_{1 2}\\
       a_{0 0} &  a_{2 1} &  a_{2 2}  
\end{array} \right],
\end{equation}
where $a_{i j}$ denotes the area covered by the drop in cell ($i j$).

The drop volume within cell ($i,j$) can be computed by 
\begin{equation}
    \centering
    \label{eq:single}
        v_{i j} = \sum_{i_m=0}^{m-1}\sum_{j_m=0}^{m-1} h_{i_m j_m} a_c
\end{equation}

Similarly, the volume of a single drop can be denoted by 
\begin{equation}
\centering
\label{eq:v_single}
    V = \left[ \begin{array}{ccc}
       v_{0 0} &  v_{0 1} &  v_{0 2}\\
       v_{0 0} &  v_{1 1} &  v_{1 2}\\
       v_{0 0} &  v_{2 1} &  v_{2 2}  
\end{array} \right],
\end{equation}
where $v_{i j}$ denotes the drop volume within cell ($i, j$).

Given $a_{i j}$ and $v_{i j}$, the average height of cell ($i,j$) denoted by $h_{i j}$ can be obtained by 
\begin{equation}
    \centering
    h_{i j} = \frac{v_{i j}}{a_{i j}} = \frac{\sum_{i_m=0}^{m-1}\sum_{j_m=0}^{m-1} h_{i_m j_m}}{M}
\end{equation}



\label{sec:equip}
Experiments were conducted to obtain the height profile model for a single drop.
The experimental setup includes a Microdrop piezo-electrical dispensing system MD-K-140, a XY Stage, a UV light and a Zeta-20 optical profiler.
The dispenser head has a heated nozzle tip of 70 $\mu m$.
Zeta-20 has a 0.35x coupler with $2/3''$ camera and we used 100x objective lens for all measurements.
The Z resolution in this setting is 0.04 $\mu m$.

Microscope slide is chosen as the substrate for its flatness.
The dispensing head is mounted on a linear stage. 
The UV inks (C-Flex 12-050 Black) are obtained from Kao Collins Inc.
Its viscosity is 17.2 $cP$ at 22.4 $\degree C$.
In this study, the pitch is set to 70 $\mu m$ and the measured drop radius $r$ is around 140 $\mu m$.

\begin{figure}
    \centering
    \includegraphics[width=\linewidth]{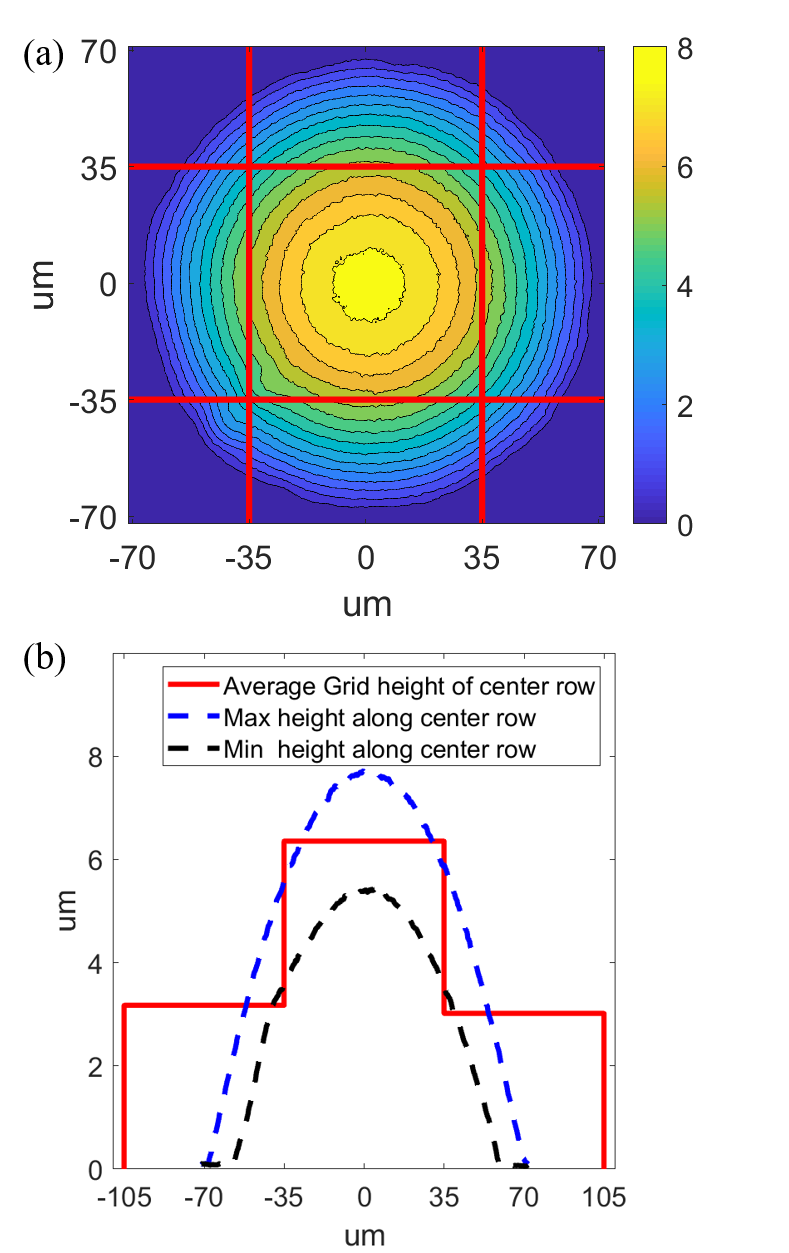}
    \caption{(a): contour of single drop, marked with 70 $\mu m$ pitch size. (b): dash lines are height measurements of the sample in a), with blue dash line showing the maximum height along the center row and black dash line showing the minimum height along the center row; red solid line shows the average cell height long center row.}
    \label{fig:s_contour}
\end{figure}

To obtain the height profile for a single drop,  50 drops are deposited on to a glass substrate.
For each drop, $a_{i j}$ and $v_{i j}$ are obtained from experimental data and Eq. (\ref{eq:single}), respectively.
The average area and volume covered by a single drop is computed as the average of the 50 samples, see Table \ref{tab:v_1} and \ref{tab:a_1}, respectively.
The height profile is then computed from the ratio between the average volume and the average area coverage, see Table \ref{tab:h_1}.
Figure \ref{fig:s_contour}.(a) is the height contour of one of the samples. 
Red lines mark the cell boundaries. 
Figure \ref{fig:s_contour}.(b) compares the measured height along center row of this sample and the average cell height along center row of 50 samples.

\begin{table}[]
	\centering
	\caption{Volume ($\mu m ^3$) in each cell for a single drop.}
	\label{tab:v_1}
	\begin{tabular}{|c|c|c|}
		\hline
		474.52 & 5480.10 & 490.56 \\ \hline
		6105.06 & 31527.32 & 6131.28 \\ \hline
		479.55 & 5646.20 & 484.74 \\ \hline 
	\end{tabular}
\end{table}

\begin{table}[]
	\centering
	\caption{Occupied area ($\mu m^2$) in each cell for a single drop.}
	\label{tab:a_1}
	\begin{tabular}{|c|c|c|}
		\hline
		340.04 & 1942.06 & 334.10 \\ \hline
		1923.37 & 4961.90 & 2033.16 \\ \hline
		338.93 & 1960.20 & 333.75 \\ \hline 
	\end{tabular}
\end{table}

\begin{table}[]
	\centering
	\caption{Average cell height ($\mu m$) in each cell for a single drop.}
	\label{tab:h_1}
	\begin{tabular}{|c|c|c|}
		\hline
		1.40 & 2.82 & 1.47 \\ \hline
		3.17 & 6.35 & 3.02 \\ \hline
		1.41 & 2.88 & 1.45 \\ \hline
	\end{tabular}
\end{table}

\section{\MakeUppercase{Height Profile Model for Multiple Adjacent Drops}}
\label{sec:model}

\begin{figure}
	\centering
	\includegraphics[width=\linewidth]{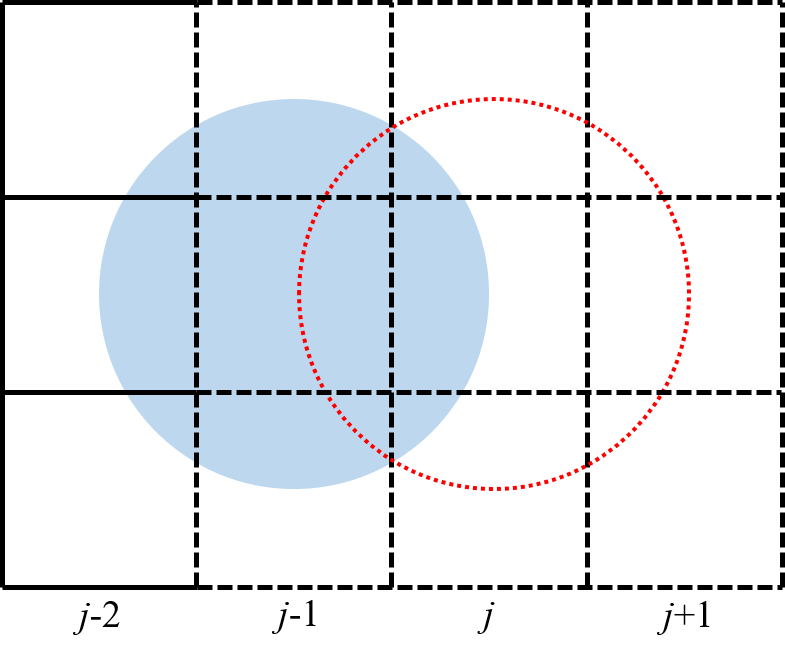}
	\caption{Light blue circle represents the $(k-1)^{th}$ drop centered at cell ($1,j-1$) and red dashed circle represents the $k^{th}$ drop centered at cell ($1,j$). When the $k^{th}$ drop is deposited, the dashed $3 \times 3$ cells are the new area of interest.}
	\label{fig:two_drop_dash}
\end{figure}

We assume non-adjacent drops deposited on non-porous flat surface have the same height profile.
When the drop is deposited on non-flat surface, localized ink flow will occur.
Such flow depends on many parameters, such as viscosity and surface tension, which requires complicated and detailed material model.
To simplify computation, we assume  that the average height in each cell changes due to additional ink volume and changes in area coverage, i.e. 
\begin{equation}
    \centering
    \label{eq:hnew}
    h_{i j,k} = \frac{v_{i j, k-1}+\Delta v_{i j}}{a_{i j, k-1}+\Delta a_{i j}},
\end{equation}
where $h_{i j,k}$ is the height after the $k^{th}$ drop, $v_{i j,k-1}$ and $a_{i j,k-1}$ are the volume and area before the $k^{th}$ drop, $\Delta v_{i j}$ and $\Delta a_{i j}$ are the changes in volume and area due to the additional drop. 

Figure \ref{fig:two_drop_dash} illustrates the impact of a new $k^{th}$ drop  deposited at cell ($1,j$), next to a previously deposited $(k-1)^{th}$ drop at cell ($1,j-1$). 
The $3 \times 3$ dashed cells represent the new area of interest associated with the new drop.

Since there can be either 0 or 1 drop in a single cell, we extend the height profile matrix $H$ notation in Eq. (\ref{eq:h_k}) to accommodate for this.
The height profile of the new area of interest centered at cell ($1,j$) is given by
\begin{equation}
    \centering
    \label{eq:h_ext}
    H_j[0] = \left[ \begin{array}{ccc}
       h_{0 j-1, k-1} &  h_{0 j, k-1} &  0  \\
       h_{1 j-1, k-1} &  h_{1 j, k-1} &  0  \\
       h_{2 j-1, k-1} &  h_{2 j, k-1} &  0  
\end{array} \right],
    H_j[1] = \left[ \begin{array}{ccc}
       h_{0 j-1, k} &  h_{0 j, k} &  h_{0 j+1, k}  \\
       h_{1 j-1, k} &  h_{1 j, k} &  h_{1 j+1, k}  \\
       h_{2 j-1, k} &  h_{2 j, k} &  h_{2 j+1, k}  
\end{array} \right],
\end{equation}
where $H_j[0]$ represents the height profile of the new area of interest before  the $k^{th}$ drop  has been deposited at cell ($1,j$); $H_j[1]$ represents the height profile of the new area of interest after the $k^{th}$ drop  has been deposited at cell ($1,j$).

Similarly, 
The drop volume of the new area of interest centered at cell ($1,j$) is given by
\begin{equation}
    \centering
    \label{eq:v_k}
    V_j[0] = \left[ \begin{array}{ccc}
       v_{0 j-1, k-1} &  v_{0 j, k-1} &  0  \\
       v_{1 j-1, k-1} &  v_{1 j, k-1} &  0  \\
       v_{2 j-1, k-1} &  v_{2 j, k-1} &  0  
\end{array} \right],
    V_j[1] = \left[ \begin{array}{ccc}
       v_{0 j-1, k} &  v_{0 j, k} &  v_{0 j+1, k}  \\
       v_{1 j-1, k} &  v_{1 j, k} &  v_{1 j+1, k}  \\
       v_{2 j-1, k} &  v_{2 j, k} &  v_{2 j+1, k}  
\end{array} \right],
\end{equation}
where $V_j[0]$ represents the drop volume in the new area of interest before  the $k^{th}$ drop  has been deposited at cell ($1,j$); $V_j[1]$ represents the drop volume in the new area of interest after the $k^{th}$ drop  has been deposited at cell ($1,j$).

The drop area coverage of the new area of interest centered at cell ($1,j$) is given by
\begin{equation}
    \centering
    \label{eq:a_k}
    A_j[0] = \left[ \begin{array}{ccc}
       a_{0 j-1, k-1} &  a_{0 j, k-1} &  0  \\
       a_{1 j-1, k-1} &  a_{1 j, k-1} &  0  \\
       a_{2 j-1, k-1} &  a_{2 j, k-1} &  0  
\end{array} \right],
    A_j[1] = \left[ \begin{array}{ccc}
       a_{0 j-1, k} &  a_{0 j, k} &  a_{0 j+1, k}  \\
       a_{1 j-1, k} &  a_{1 j, k} &  a_{1 j+1, k}  \\
       a_{2 j-1, k} &  a_{2 j, k} &  a_{2 j+1, k}  
\end{array} \right],
\end{equation}
where $A_j[0]$ represents the drop area coverage in the new area of interest before  the $k^{th}$ drop  has been deposited at cell ($1,j$); $A_j[1]$ represents the drop area coverage in the new area of interest after the $k^{th}$ drop  has been deposited at cell ($1,j$).

From Eq. (\ref{eq:v_k}) and (\ref{eq:a_k}), assuming a first drop has been deposited at cell (1,1), the drop volume and area coverage in the new  area of interest ($3 \times 3$ area centered at cell (1,2)) before a second drop depositing at cell (1,2) is 
\begin{equation}
    \label{eq:h_2}
    V_2[0] = \left[ \begin{array}{ccc}
       v_{0 1, 1} &  v_{0 2, 1} &  0  \\
       v_{1 1, 1} &  v_{1 2, 1} &  0  \\
       v_{2 1, 1} &  v_{2 2, 1} &  0  
\end{array} \right], \text{\hspace{12pt} and \hspace{12pt} }
A_2[0] = \left[ \begin{array}{ccc}
       a_{0 1, 1} &  a_{0 2, 1} &  0  \\
       a_{1 1, 1} &  a_{1 2, 1} &  0  \\
       a_{2 1, 1} &  a_{2 2, 1} &  0  
\end{array} \right]
\end{equation}
where the first two columns of $V_2[0]$ and $A_2[0]$ are the second and third columns of $V_1[1]$ and $A_1[1]$, where
\begin{equation}
    V_1[1] = \left[ \begin{array}{ccc}
       v_{0 0, 1} &  v_{0 1, 1} &  v_{0 2, 1}  \\
       v_{1 0, 1} &  v_{1 1, 1} &  v_{1 2, 1}  \\
       v_{2 0, 1} &  v_{2 1, 1} &  v_{2 2, 1}  
\end{array} \right], \text{\hspace{12pt} and \hspace{12pt} }
A_1[1] = \left[ \begin{array}{ccc}
       a_{0 0, 1} &  a_{0 1, 1} &  a_{0 2, 1}  \\
       a_{1 0, 1} &  a_{1 1, 1} &  a_{1 2, 1}  \\
       a_{2 0, 1} &  a_{2 1, 1} &  a_{2 2, 1}  
\end{array} \right].
\end{equation}


Using the above notation, a height profile propagation model can be written as 
\begin{equation}
    \centering
    \label{eq:h_prop}
      V_j[1] = V_j[0] + \Delta V \text{ \hspace{0.03\textwidth}and\hspace{0.03\textwidth} }
      A_j[1] = A_j[0] + \Delta A,
\end{equation}
where $\Delta V, \Delta A$ are cell volume change matrix and area coverage change matrix, respectively.
Given the volume and area associated with the new area of interest, the height profile for the new area of interest is 
\begin{equation}
\label{eq:h_jk}
    H_j[1] = \frac{V_j[1]}{A_j[1]},
\end{equation}
where the division above is elemental division.

For the two drop example above, 
after the second drop has been deposited at cell (1,2), the height profile in the new area of interest is 
\begin{equation}
    H_2[1] = \left[ \begin{array}{ccc}
       h_{0 1, 2} &  h_{0 2, 2} &  h_{0 3, 2}  \\
       h_{1 1, 2} &  h_{1 2, 2} &  h_{1 3, 2}  \\
       h_{2 1, 2} &  h_{2 2, 2} &  h_{2 3, 2}  
\end{array} \right] = \left[\frac{v_{ij2}}{a_{ij2}}\right],
\end{equation}
where $i =0,1,2$ and $j=1,2,3$.



This section is further divided into 2 parts, with cell volume propagation model in Section \ref{sec:vol_model} and area coverage propagation model in Section \ref{sec:area_model}.

\subsection{\MakeUppercase{Volume Propagation}}
\label{sec:vol_model}


Experimental data will be used to determine $\Delta V$ for the volume propagation model. 
The single drop volume is given by Eq. (\ref{eq:v_single}), i.e.
\begin{equation}
    V = \left[ \begin{array}{ccc}
       v_{0 0} &  v_{0 1} &  v_{0 2}\\
       v_{0 0} &  v_{1 1} &  v_{1 2}\\
       v_{0 0} &  v_{2 1} &  v_{2 2}  
\end{array} \right] = V_1[1] = \left[ \begin{array}{ccc}
       v_{0 0, 1} &  v_{0 1, 1} &  v_{0 2, 1}  \\
       v_{1 0, 1} &  v_{1 1, 1} &  v_{1 2, 1}  \\
       v_{2 0, 1} &  v_{2 1, 1} &  v_{2 2, 1}  
\end{array} \right] 
.
\end{equation}
$V_2[0]$ can be written as 
\begin{equation}
       V_2[0] = \left[ \begin{array}{ccc}
       v_{0 1, 1} &  v_{0 2, 1} &  0\\
       v_{1 1, 1} &  v_{1 2, 1} &  0\\
       v_{2 1, 1} &  v_{2 2, 1} &  0
\end{array} \right].
\end{equation}
The second drop is deposited at cell ($1,2$) after the drop at cell (1,1) is cured with UV light.
After the second drop is cured, $V_2[1]$ can be obtained from experiment. 
$\Delta V$ can then be obtained by 
\begin{equation}
\label{eq:cal_dv}
    \Delta V = V_2[1] - V_2[0] , \text{ or }
    \Delta V = [\Delta v_{ij}] = \left[v_{i j, 2}-v_{i j, 1} \right].
\end{equation}

\subsection{\MakeUppercase{Area Propagation}}
\label{sec:area_model}

Area propagation follows the same approach as volume propagation, where the single drop area coverage is given by Eq. (\ref{eq:a_single}), i.e.
\begin{equation}
    A = \left[ \begin{array}{ccc}
       a_{0 0} &  a_{0 1} &  a_{0 2}\\
       a_{0 0} &  a_{1 1} &  a_{1 2}\\
       a_{0 0} &  a_{2 1} &  a_{2 2}  
\end{array} \right] = A_1[1] = \left[ \begin{array}{ccc}
       a_{0 0, 1} &  a_{0 1, 1} &  a_{0 2, 1}  \\
       a_{1 0, 1} &  a_{1 1, 1} &  a_{1 2, 1}  \\
       a_{2 0, 1} &  a_{2 1, 1} &  a_{2 2, 1}  
\end{array} \right] 
.
\end{equation}
$A_2[0]$ can be written as 
\begin{equation}
       A_2[0] = \left[ \begin{array}{ccc}
       a_{0 1, 1} &  a_{0 2, 1} &  0\\
       a_{1 1, 1} &  a_{1 2, 1} &  0\\
       a_{2 1, 1} &  a_{2 2, 1} &  0
\end{array} \right].
\end{equation}
The second drop is deposited at cell ($1,2$) after the drop at cell (1,1) is cured with UV light.
After the second drop is cured, $A_2[1]$ can be obtained from measurement. 
$\Delta A$ can then be obtained by 
\begin{equation}
\label{eq:cal_da}
    \Delta A = A_2[1] - A_2[0] , \text{ or }
    \Delta A = [\Delta a_{ij}] = \left[a_{i j, 2}-a_{i j, 1} \right].
\end{equation}

\section{\MakeUppercase{Experimental Validation}}
\label{sec:exp_valid}
Experiments are carried out to validate the propagation model.
We calculate $\Delta V$ and $\Delta A$ from 2-drop experiments.
Then we use them to predict the height profile from a  3-drop line ($H_3[1]$).
Analysis is based on RMS error, which can be calculated by
\begin{equation}
    \centering
    \label{eq:rms}
    RMS = \sqrt{\frac{1}{N}\left(\frac{V_p-V_m}{V_m}\right)^2},
\end{equation}
where $N$ is number of samples; $V_p$ and $V_m$ are predicted and measured values, respectively.

To obtain enough data, we print and analyze 50 samples of single-drop, 2-drop and 3-drop pattern each, at 70 $\mu m$ pitch.
Single-drop data is shown in Section \ref{sec:single}.





\subsection{\MakeUppercase{Calculate $\Delta V$ and $\Delta A$}}
\label{sec:cal_dV_dA}
To avoid coalescence, the first drop is cured under UV light for 1.5 seconds, before the second drop is deposited.
This longer time is setup to capture dynamics of each additional drop.
The printing orders can be adjusted, such that all non-adjacent drops can be printed and cured at once. 
As a result, large format printing would not take a long time. 
With 50 samples in total, we randomly separate 2-drop samples to 3 groups of 15 samples and 1 group of 5 samples.
Two groups of 15 samples are used to obtain $V_2[1]$ and $A_2[1]$.
Then $V_2[1]$ and $A_2[1]$ are compared with samples in the other groups, to confirm consistency of results.


2-drop measurements are shown in Table \ref{tab:v_2}, \ref{tab:a_2} and \ref{tab:h_2}, with predicted values in bold.
The cell volume change matrix ($\Delta V$) and the area coverage change matrix ($\Delta A$) can be obtained from Eq. (\ref{eq:cal_dv}) and (\ref{eq:cal_da}), shown in Table \ref{tab:delta_v} and \ref{tab:delta_a}, respectively.
Two values of $\Delta A$ are not calculated, because the two cells will always be fully filled.
Values are larger in cells without prior ink, which is as expected.
\begin{table}[]
	\centering
	\caption{Measured and predicted cell volume ($\mu m ^3$) in each cell of 2 drops.}
	\label{tab:v_2}
	\begin{tabular}{|c|c|c|c|}
		\hline
		442.99 & 5277.37 & 6719.44 & 1385.46 \\ 
		(\textbf{474.52}) & (\textbf{5875.85}) & (\textbf{6962.56}) & (\textbf{1444.70}) \\ \hline
		5955.36 & 32180.14  & 35927.41  & 10632.84 \\
		(\textbf{6105.06}) & (\textbf{32162.25}) & (\textbf{35830.02}) & (\textbf{10128.87}) \\ \hline
		467.12 & 5350.54 & 7301.58 & 1343.83 \\ 
		(\textbf{4779.55}) & (\textbf{6042.79}) & (\textbf{6927.05}) & (\textbf{1406.72}) \\ \hline 
	\end{tabular}
\end{table}


\begin{table}[]
	\centering
	\caption{Measured and predicted area coverage ($\mu m ^2$) in each cell of 2 drops.}
	\label{tab:a_2}
	\begin{tabular}{|c|c|c|c|}
		\hline
		312.36 & 1908.15 & 2176.14 & 549.70 \\ 
		(\textbf{340.04}) & (\textbf{1981.17}) & (\textbf{2039.23}) & (\textbf{557.95}) \\ \hline
		1946.35 & 4984.03 & 4984.03 & 3013.14  \\ 
		(\textbf{1923.37}) & (\textbf{4961.9}) & (\textbf{4961.9}) & (\textbf{2862.69}) \\ \hline
		314.83 & 1965.58 & 2007.68 & 556.97 \\ 
		(\textbf{338.93}) & (\textbf{2011.72}) & (\textbf{2078.33}) & (\textbf{543.01}) \\ \hline 
	\end{tabular}
\end{table}


\begin{table}[]
	\centering
	\caption{Calculated $\Delta V$}
	\label{tab:delta_v}
	\begin{tabular}{|c|c|c|}
		\hline
		395.75 & 6472.00 & 1444.70 \\ \hline
		634.93 & 29698.74 & 10128.87 \\ \hline
		396.59 & 6442.31 & 1406.72\\ \hline 
	\end{tabular}
\end{table}

\begin{table}[]
	\centering
	\caption{Calculated $\Delta A$}
	\label{tab:delta_a}
	\begin{tabular}{|c|c|c|}
		\hline
		39.10 & 1705.13 & 557.95 \\ \hline
		N/A & N/A & 2862.69 \\ \hline
		51.52 & 1744.58 & 543.01 \\
		\hline 
	\end{tabular}
\end{table}

\begin{table}[]
	\centering
	\caption{Measured and predicted height ($\mu m$) in each cell of 2 drops.}
	\label{tab:h_2}
	\begin{tabular}{|c|c|c|c|}
		\hline
        1.42 & 2.77 & 3.09 & 2.52 \\ 
        (\textbf{1.40}) & (\textbf{2.97}) & (\textbf{3.41}) & (\textbf{2.59}) \\ \hline
        3.06 & 6.46 & 7.21 & 3.53 \\ 
        (\textbf{3.17}) & (\textbf{6.48}) & (\textbf{7.22}) & (\textbf{3.54}) \\ \hline
        1.48 & 2.72 & 3.64 & 2.41 \\ 
        (\textbf{1.41}) & (\textbf{3.00}) & (\textbf{3.33}) & (\textbf{2.59}) \\ \hline
	\end{tabular}
\end{table}


\subsection{\MakeUppercase{Results of 3-drop Pattern}}
\label{sec:result}
After obtaining $\Delta V$ and $\Delta A$, it is possible to check how well does the propagation model predict height profiles of 3 consecutive drops. 
So we use Eq. (\ref{eq:h_prop}) to predict $V_3[1]$ and $A_3[1]$ before obtaining predicted $H_3[1]$ with Eq. (\ref{eq:h_jk}).
Table \ref{tab:v_3}, \ref{tab:a_3} and \ref{tab:h_3} show measured cell volume, area coverage and cell height in each cell, with predicted values in bold, respectively.
Error analysis is based on RMS errors, which can be calculated with Eq. (\ref{eq:rms}).

\begin{table}[]
	\centering
	\caption{Measured and predicted volume ($\mu m ^3$) in each cell of 3 drops.}
	\label{tab:v_3}
	\begin{tabular}{|c|c|c|c|c|}
		\hline
		572.0 & 6033.9 & 7411.2 & 7183.1 & 1948.9 \\  
		(\textbf{443.0}) & (\textbf{5277.4}) & (\textbf{7115.2}) & (\textbf{7857.5}) & (\textbf{1444.7})\\ \hline
		6920.21 & 31280.0 & 36622.8 & 36940.6 & 15521.6  \\ 
		(\textbf{5955.4}) & (\textbf{32180.1}) & (\textbf{36562.3}) & (\textbf{40331.58}) & (\textbf{10128.8}) \\ \hline
		686.26 & 6673.4 & 7270.7 & 6361.44 & 1904.9 \\ 
		(\textbf{467.1}) & (\textbf{5350.5}) & (\textbf{7698.2}) & (\textbf{7786.1}) & (\textbf{1406.7}) \\ \hline 
	\end{tabular}
\end{table}


\begin{table}[]
	\centering
	\caption{Measured and predicted area ($\mu m ^2$) in each cell of 3 drops.}
	\label{tab:a_3}
	\begin{tabular}{|c|c|c|c|c|}
		\hline
		459.9 & 2325.3 & 2491.0 & 2424.3 & 833.4 \\ 
		(\textbf{312.4}) & (\textbf{1908.2}) & (\textbf{2215.2}) & (\textbf{2254.8}) & (\textbf{557.9}) \\ \hline
		2288.3 & 4956.2 & 4952.4 & 4938.8 & 3902.1 \\  
		(\textbf{1946.4}) & (\textbf{4984.0}) & (\textbf{4984.0}) & (\textbf{4984.0}) & (\textbf{2862.7}) \\ \hline
		448.6 & 2317.7 & 1955.5 & 1844.0 & 776.6 \\ 
		(\textbf{314.8}) & (\textbf{1965.6}) & (\textbf{2059.2}) & (\textbf{2301.5}) & (\textbf{543.0}) \\ \hline 
	\end{tabular}
\end{table}


\begin{table}[]
	\centering
	\caption{Measured and predicted height ($\mu m$) in each cell of 3 drops.}
	\label{tab:h_3}
	\begin{tabular}{|c|c|c|c|c|}
		\hline
		1.24 & 2.59 & 2.98 & 2.96 & 2.34 \\ 
		(\textbf{1.42}) & (\textbf{2.77}) & (\textbf{3.21}) & (\textbf{3.48}) & (\textbf{2.59}) \\ \hline
		3.02 & 6.31 & 7.39 & 7.48 & 3.98 \\  
		(\textbf{3.06}) & (\textbf{6.46}) & (\textbf{7.34}) & (\textbf{8.09}) & (\textbf{3.54}) \\ \hline
		1.53 & 2.66 & 3.72 & 3.45 & 2.45\\ 
		(\textbf{1.48}) & (\textbf{2.72}) & (\textbf{3.74}) & (\textbf{3.38}) & (\textbf{2.59}) \\ \hline 
	\end{tabular}
\end{table}

 
\begin{figure}
	\centering
	\includegraphics[width=\linewidth]{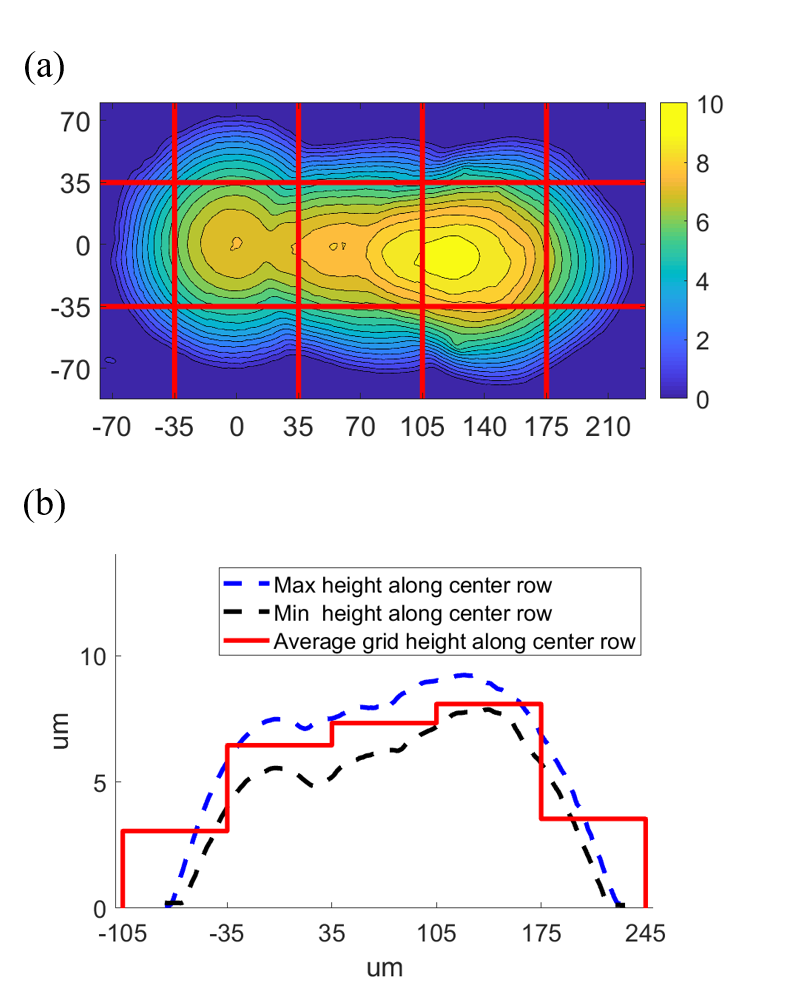}
	\caption{(a): Contour of 3 drops, marked with 70 $\mu m$ pitch size. \hspace{\textwidth}(b): Dash lines are height measurements of the sample in a), with blue dash line showing the maximum height along the center row and black dash line showing the minimum height along the center row; red solid line shows the average cell height long center row.}
	\label{fig:3_comp}
\end{figure}

Figure \ref{fig:3_comp}.(a) is the height contour of one of the 3-drop samples, with red solid line marking the boundaries of each cell.
Figure \ref{fig:3_comp}.(b) shows the maximum and minimum height of each column measured along center row, and predicted height profile.
It can be observed that the predicted height match the trend of actual measurements.
RMS error is 6.6\% along the center row, with 7.4\% among all cells.
Table \ref{tab:comp} shows the comparison of modeling RMS error among 3 height models. 
Both graph-based dynamic model and superposition model are above 10\%.
\begin{table}[]
    \centering
    \caption{RMS error comparison among 3 models}
    \label{tab:comp}
    \begin{tabular}{|c|c|c|c|}
        \hline
         & Superposition & Graph-based & This Model \\ \hline
         RMS Error & 14.2\% & 17.2\% & 7.4\% \\ \hline
    \end{tabular}
\end{table}

\subsection{\MakeUppercase{Height Profile Stability Test}}
\label{sec:stability}
3-drop prediction confirms that our proposed model works better than existing models.
Cell height along center row shows an upward trend, which raises concern on stability. 
From common sense, we know that printing a line will not lead to monotonically increasing cell height along center row.
The predicted maximum cell height at center row can be calculated by
\begin{equation}
    \label{eq:pk_height}
    \centering
    h_{\text{max}} = \frac{\sum_{j=0}^2 \Delta v_{1j}}{a_{11}} = 8.15 \mu m.
\end{equation}

Besides theoretical calculation, experimental results also confirm the stability of height profiles.
Table \ref{tab:max_height} shows the peak cell height from 3-drop, 4-drop and 5-drop printings. 
The maximum deviation from predicted max cell height is $8.2 \%$. 

\begin{table}[]
    \centering
    \caption{Max cell height of 3-drop, 4-drop and 5-drop printing.}
    \label{tab:max_height}
    \begin{tabular}{|c|c|c|c|}
    \hline
    Number of Drops & 3 & 4 & 5 \\ \hline
    Max Cell Height ($\mu m$) & 7.48 & 8.19 & 8.02 \\ \hline
    \end{tabular}
\end{table}

\section{\MakeUppercase{Conclusion}}
\label{sec:conclusion}
In this paper, we proposed a height profile propagation model for drop-on-demand printing of UV curable inks.
The model uses one and two drop height profiles to predict the height profiles of subsequent drops.
The fundamental idea is based on material flow due to substrate height difference.
3-drop height profile prediction is validated with experiments.
RMS error along center row is 6.2\% and that of all cells is 7.4\%, both of which are better than existing models.
Additional work is needed to analyze the stability of the proposed propagation model as well as additional experimental validation are needed.
\bibliographystyle{plain}
\bibliography{library.bib}

\begin{thebibliography}{10}

\bibitem{Agar2005}
A.~Ufuk Agar and Jan~P. Allebach.
\newblock {Model-based color halftoning using direct binary search}.
\newblock {\em IEEE Transactions on Image Processing}, 14(12):1945--1959, 2005.

\bibitem{standard2012iso}
{ASTM International}.
\newblock {ISO/ASTM}52900-15 standard terminology for additive manufacturing
  – general principles – terminology.
\newblock {\em ASTM International}, 2015.

\bibitem{choi2017numerical}
Moonhyeok Choi, Gihun Son, and Woosup Shim.
\newblock Numerical simulation of droplet impact and evaporation on a porous
  surface.
\newblock {\em International Communications in Heat and Mass Transfer},
  80:18--29, 2017.

\bibitem{cooley2002applicatons}
Patrick Cooley, David Wallace, and Bogdan Antohe.
\newblock Applicatons of ink-jet printing technology to biomems and
  microfluidic systems.
\newblock {\em JALA: Journal of the Association for Laboratory Automation},
  7(5):33--39, 2002.

\bibitem{Doumanidis2000}
Charalabos Doumanidis and Eleni Skordeli.
\newblock {Distributed-Parameter Modeling for Geometry Control of Manufacturing
  Processes With Material Deposition}.
\newblock {\em Journal of Dynamic Systems, Measurement, and Control},
  122(1):71, 2000.

\bibitem{ford2016additive}
Simon Ford and M{\'e}lanie Despeisse.
\newblock Additive manufacturing and sustainability: an exploratory study of
  the advantages and challenges.
\newblock {\em Journal of Cleaner Production}, 137:1573--1587, 2016.

\bibitem{gunjal2003experimental}
PR~Gunjal, VV~Ranade, and RV~Chaudhari.
\newblock Experimental and computational study of liquid drop over flat and
  spherical surfaces.
\newblock {\em Catalysis today}, 79:267--273, 2003.

\bibitem{Guo2018}
Yijie Guo, Joost Peters, Tom Oomen, and Sandipan Mishra.
\newblock {Control-oriented models for ink-jet 3D printing}.
\newblock {\em Mechatronics}, 56:211--219, 2018.

\bibitem{le1998progress}
Hue~P Le.
\newblock Progress and trends in ink-jet printing technology.
\newblock {\em Journal of Imaging Science and Technology}, 42(1):49--62, 1998.

\bibitem{Li2004}
Pingshan Li and Jan~P. Allebach.
\newblock {Tone-dependent error diffusion}.
\newblock {\em IEEE Transactions on Image Processing}, 13(2):201--215, 2004.

\bibitem{sirringhaus2003inkjet}
Henning Sirringhaus and Tatsuya Shimoda.
\newblock Inkjet printing of functional materials.
\newblock {\em MRS bulletin}, 28(11):802--806, 2003.

\end{thebibliography}



\end{document}